\documentclass[12pt]{iopart}
\usepackage{iopams}  
\usepackage{graphicx}
\usepackage{xspace}
\usepackage{booktabs}
\usepackage{cite}
\usepackage{lscape}
\newcommand{\hethree}{\textsuperscript{3}He\xspace}
\newcommand{\hefive}{\textsuperscript{5}He\xspace}
\newcommand{\lifive}{\textsuperscript{5}Li\xspace}
\newcommand{\hesix}{\textsuperscript{6}He\xspace}
\newcommand{\lisix}{\textsuperscript{6}Li\xspace}
\newcommand{\liseven}{\textsuperscript{7}Li\xspace}
\newcommand{\besix}{\textsuperscript{6}Be\xspace}
\newcommand{\beseven}{\textsuperscript{7}Be\xspace}
\newcommand{\benine}{\textsuperscript{9}Be\xspace}
\newcommand{\beeight}{\textsuperscript{8}Be\xspace}

\begin{document}

\title[
Reactions induced by 30 MeV \hethree beam on \benine: Cluster-transfer reactions
]
{
Reactions induced by 30 MeV \hethree beam on \benine: Cluster transfer reactions
}

\author{Urazbekov~B~A$^{1,2,3}$, Issatayev~T$^{1,2,4*}$, Lukyanov~S~M$^{4}$, Azhibekov~A$^{2,4,5}$,  Denikin A~S$^{4,6}$, Mendibayev~K$^{2,4}$, Janseitov~D~M$^{2,3,7}$, Penionzhkevich~Yu~E$^{4}$, Kuterbekov~K~A$^{1}$, Zholdybayev~T~K$^{2,7}$}

\address{$^1$Gumiliyev Eurasian National University,  2 Satpayev Str,Astana, Kazakhstan}
\address{$^2$Institute of Nuclear Physics, 1 Ibragimov Str, Almaty, Kazakhstan}
\address{$^3$Bogolubov Laboratory of Theoretical Physics, JINR, 20 Joliot Curie Str, Dubna, Russia}
\address{$^4$Flerov Laboratory of Nuclear Reactions, JINR, 20 Joliot Curie Str, Dubna, Russia}
\address{$^5$Korkyt-ata State University,  29A Aiteke-Byi Str, Kyzylorda, Kazakhstan}
\address{$^6$Dubna State University,  19 Universitetskaya Str, Dubna, Russia}
\address{$^7$Al-Farabi Kazakh National University, Almaty, Kazakhstan}
\ead{$^*$issatayev@jinr.ru}
\vspace{10pt}
\begin{indented}
\item[]
\end{indented}

\begin{abstract}
An experiment has been carried out for studying the cluster structure of $^9$Be induced by the $^3$He ions at the energy of 30 MeV. As results of the nuclear reaction \hethree + \benine the differential cross sections for the exit channels -- elastic, inelastic, $\alpha$ + \beeight,  \hesix + \besix, \lisix + \lisix, and  \beseven + \hefive~ --  were measured.

Elastic and inelastic scattering data are treated within both the optical model and Coupled channels method.
A new set of optical potential was taken for the elastic scattering.
The deformation parameter $\delta_2$ was established for the transition $3/2\rightarrow5/2$.

Cluster transfer reactions are analyzed by means of the coupled reaction channels method. 
The nuclear reactions with the exit channels \hesix + \besix, \lisix + \lisix, and \beseven + \hefive~  are complemented by two-step transfer mechanisms. 
The contribution of each reaction mechanisms are shown, and compared with the findings of other authors.

\end{abstract}

%
\vspace{2pc}

\footnotesize

\maketitle
%
%

\section*{Introduction}
In light nuclei, e.g. with $A$ $<$ 12, in most cases the nucleons may form a group. The group in nuclear physics is often called the cluster.
Their relative motion determines general characteristics and properties. 
In this manner, studying the cluster structure of the light nuclei has become one of the prior task of nuclear physics with regard to the theory and experimental explorations. 

The nucleus \benine is an attractive nucleus due to its many internal properties. 
Low binding energy of the $p$-shell neutron $-$1.66 MeV (e.g., see \cite{nrv}), large quadrupole moment +53.3$\pm 3$ mb \cite{lederer1978table, ajzenberg1988energy}, and positive parity in the first excited state $\frac{1}{2}^+$ at 1.684 MeV (e.g., see \cite{nrv})  -- all of these experimental data points out its unique structure. Therefore, in the cluster model framework the nucleus can be supposed to have the configurations, such as $2\alpha+n$, $\alpha+^5{\rm He}$, \beeight + $n$, and other.

The cluster structure of \benine has been extensively studied within the various approaches \cite{lichtenthaler2021experiments, huang2021dipole, fortune2021methods, starastsin2021structures,maridi2021p+, mahmoud2021microscopic}.
In recent studies from Refs. \cite{amar2022analysis, amar2021elastic}, the elastic scattering data of $\alpha$-particles and of $d$ on \benine have been treated within the cluster models $\alpha+^5\rm{He}$, and \beeight$+n$. 
Authors showed that on the example of the elastic scattering, the interaction potentials of projectiles  can be treated by means of the cluster folding model within the configurations  $\alpha+^5\rm{He}$ and \beeight$+n$. 
Moreover, at the backward angles the spectroscopic amplitudes were extracted in the calculations of the elastic transfer reactions.

It is interesting to note an analysis of the two-neutron transfer in the nuclear reactions induced by \beseven radioactive beam on the target nucleus \benine \cite{umbelino2019two}. 
The two-nucleon spectroscopic amplitudes were obtained within the shell model framework and used for the elastic transfer channel.
Authors could explain the experimental data of excess cross sections by means of the CRC calculations combining both elastic scattering and elastic transfer mechanisms.
Consequently, this draws attention to that the nucleus \benine may have also a cluster structure like \beseven + $2n$.

In the study \cite{rudchik1996strong}  the nuclear reactions resulting from  $^3$He + \benine at the energy of 63 MeV have been analyzed. 
The elastic channel as well as the exit channels \beseven + \hefive, \lisix + \lisix were calculated within the Coupled Reaction Channels (CRC) method. 
For these channels, all the possible transfer mechanisms were supposed. 
In particular, the two-step transfer mechanisms beginning with $n$ pick-up well enhances the underestimated cross sections for the  channels   \beseven + \hefive and \lisix + \lisix. 
It shows that the $p$-shell neutron is loosely bound, and, in turn, the nucleus \benine has the cluster structure \beeight$+ n$. 

The same experimental study with the nuclear reaction \hethree + \benine was performed, but at the laboratory energy 30 MeV \cite{lukyanov2016cluster}. 
Pure optical model and DWBA calculations performed in this work require more sophisticated theoretical models.
Therefore, our aim is to analyze the experimental data \cite{lukyanov2016cluster} within the cluster model of \benine and to probe all possible transfer mechanisms. We focus on the internal structure of \benine presenting fresh data missing in Ref. \cite{lukyanov2016cluster, urazbekov2023three}, and on the data in more accurate processed form.

Current work is organized as follows. 
Detailed information of the experimental method are given in the first section. 
Second section is dedicated for the elastic scattering together with inelastic channel. 
The next section is for the cluster transfer channels.
Main findings with conclusion are drawn in the end.

\section{Experimental procedure}
The experiment on the nuclear reaction \hethree and \benine at the laboratory energy of 30 MeV was conducted at the Nuclear Physics Institute (NPI), \v{R}e\v{z}, Czech Republic. In the course of the experiment time, an average beam current was kept to 10 enA. The self-supporting \benine target having the thickness of  11 $\mu\rm{m}$   was developed with the foil highly purified up to 99\%. The resonances of the contamination by the carbon and oxygen isotopes were not detected in the energy spectra.

Particle identification was based on the $\Delta E$-$E$ method, i.e. measurements of the energy-loss $\Delta E$ and the residual energy $E_r$.
Four Si-Si(Li) telescopes consisting of the detectors $\Delta E_0$, $\Delta E$, and $E_r$ were mounted to register scattered ions with the thicknesses of 10$\mu$m, 100 $\mu$m, and 3 mm, respectively. The acceptance angle of each telescope was $\sim$ $0.6^\circ$ in the scattering plane, and theirs solid angle was $\sim$ 0.2 msr. The good energy resolutions of both $\Delta E$ and $E_r$ detectors provided  unambiguous identification of $A$ and $Z$ of each product. The over-all energy resolution was 150 -- 200 keV.   
 
\begin{figure}
\centering
\includegraphics[scale=0.245]{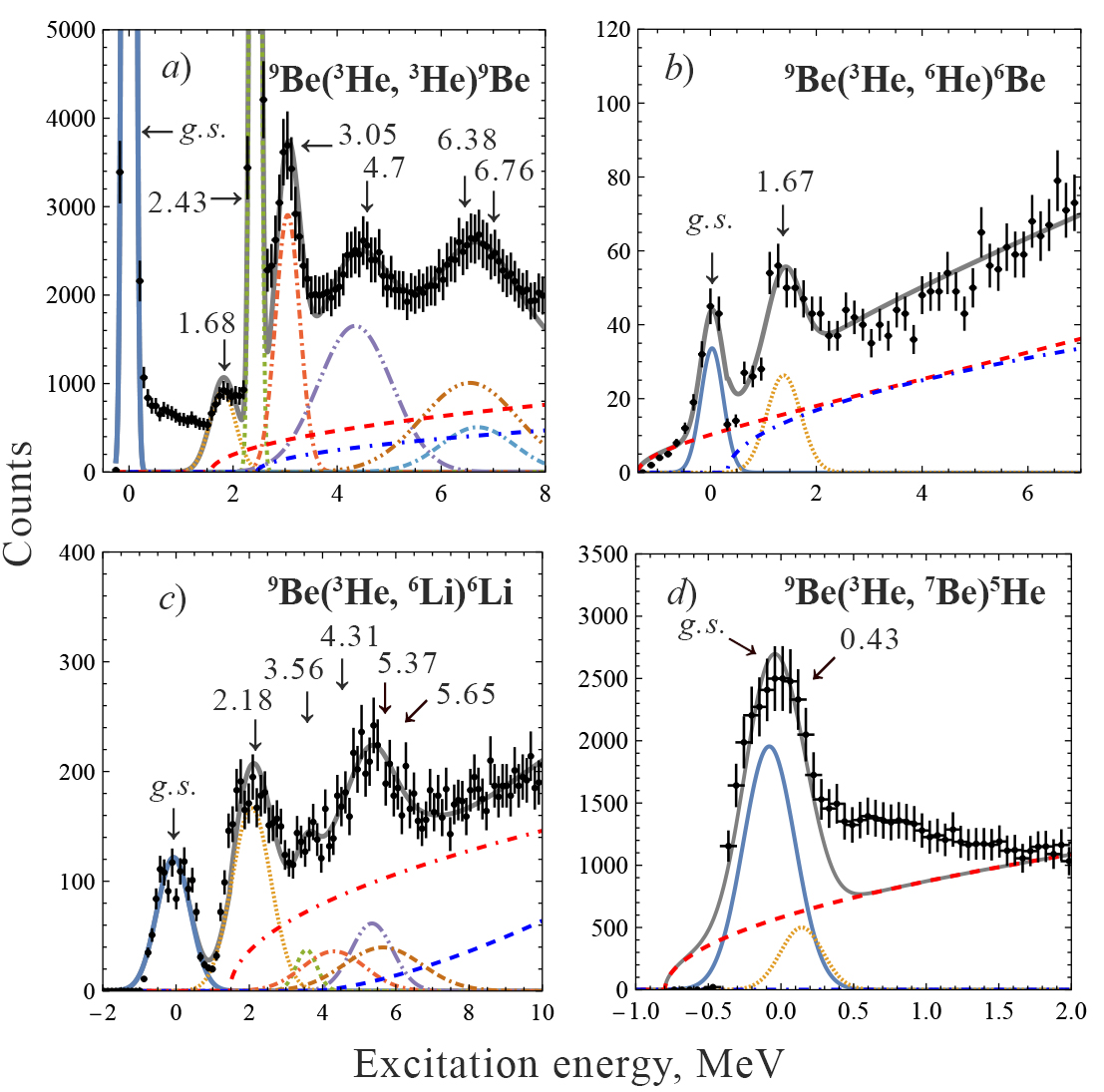}
\caption{   Excitation energy spectra of $^9$Be (a), $^6$Be (b), $^6$Li (c) and $^5$He (d) for the nuclear reaction   $^3$He $+$ $^9$Be at $E_{lab}=30$ MeV measured at $\theta_{lab}$ = $12^\circ$. The total fit for each spectrum and subtracted peaks are shown. The background was fitted taking into account the three-body and four-body breakup processes (For more, please, see the text).
}
\label{experiment}
\end{figure}

 
The excitation energy spectra for \benine, \besix, \lisix and \hefive are shown in Fig.\ref{experiment}. The ground and excited states of the nuclei \benine, \besix, \lisix and \hefive were identified in the following reaction channels: \benine(\hethree,~\hethree)\benine (see panel \textit{a}), \benine(\hethree, \hesix)\besix (\textit{b}), \benine(\hethree, \lisix)\lisix (\textit{c}) and \benine(\hethree, \beseven)\hefive (\textit{d}).
The energy calibration of the $\Delta Es$ and $E_r$ detectors has been performed 
taking into account well-known states which are strongly excited in the spectra. 
The calibration appeared to be practically linear. 
This allows us to determine the positions of all excited states in the spectra. 
Total energies were calculated as the sum over the calibrated energy losses $\Delta E_s$ and the residual energy $E_r$.
The excitation energy spectra were constructed as $E_{g.s}$ (position) - $E_{total}$.

The peaks identified by means of the fitting procedure adopting the method of standard Gaussian decomposition.
With the known energy of calibration, the positions of peaks and widths were fixed in accordance with the generally accepted experimental data.
It should be noted that the width of the state in each spectrum can include: the natural width, the apparatus width of set-up and the energy spread.
The latter case is originated by 
the emitting particles, which their production can occur either at the beginning or at the end of the target foil.
The background areas are also illustrated in Fig.\ref{experiment}. They are mainly due to the phase volumes which depend on the threshold energy in the following break-up processes: 
\begin{itemize}
 \item \benine $\rightarrow$ $^8\rm{Be}$ + $n$ (see Fig. \ref{experiment}, panel \textit{a}, red dashed curve) or $^4\rm{He}$ + $\rm{\hefive}$ (\textit{a}, blue dot-dashed);
 \item\besix $\rightarrow$ $^4\rm{He}$ + 2$p$ (\textit{b}, red dashed) or $^5\rm{Li}$ + $p$ (\textit{b}, blue dot-dashed); 
 \item\lisix $\rightarrow$ $^4\rm{He}$ + $d$ (\textit{c}, red dot-dashed) or $^4\rm{He}$ + $p$ + $n$ (\textit{c}, blue dashed); 
 \item\hefive $\rightarrow$ $^4\rm{He}$ + $n$ (\textit{d}, red dashed).
\end{itemize}


The absolute error in the determination of cross section is not more than 15\%. 
Consequently, the following components can mostly contribute to the error: in the decomposition, statistical error of the number of events in the subtracted peak,  error in determining the thickness of target and in the values of solid angle, loss-events, current measurement error.

\section{Elastic and inelastic channels}
\subsection*{Elastic scattering}

The differential cross section of the elastic scattering of \hethree from the nucleus \benine was treated within the optical model framework. 
Numerical calculations were carried out by means of the \texttt{FRESCO} code \cite{thompson1988coupled}. 
The optical potential used in the OM calculations was taken in the form:

\begin{equation}\label{eqn:OP}
\begin{array}{l}
 U(R)=-V(R)-iW(R)+V^{SO}(R)(\textbf{l} \cdot \sigma)+V^C(R),
\end{array}
\end{equation}
where $R$ is the distance between the \hethree  and \benine, and $V^{V}, W^{V}$ are the real and imaginary volume potential terms, $ V^{SO}$ and $V^C$ are the spin-orbit and Coulomb potentials. The volume potentials may represent the Woods-Saxon (WS) potential \cite{woods1954diffuse}:
\begin{equation}
V\left( R \right) =  V_0^V f_{R_V, a_V} \left( R \right),~~~~f_{R_0, a_0} \left( R \right) =   \left[ 1+ \exp \left( \frac{R-R_0}{a_0} \right) \right]^{-1},
\end{equation}
where $V_0$ is the depth of the potential, $r_0$ is  the average distance, and $a_0$ is the  diffusion parameter. 
The spin-orbit  term can have the form as follows:
\begin{eqnarray}
V^{SO}(R) &= V_0^{SO}\left(\frac{\hbar}{m_\pi c}\right)^2 \frac{1}{R} \frac{d}{dR} f_{R_{SO} a_{SO}}(R),
\end{eqnarray}
while the Coulomb term was taken as the interaction of a point-charge with a uniformly charged sphere
\begin{eqnarray}
\label{coul}
V^C(R)=
\cases{
\frac{Z_1 Z_2 e^2}{2 R_C} \left( 3- \frac{R^2}{R_C^2} \right), & for  R $\leq R_C$, \\
\frac{Z_1 Z_2 e^2}{R}, & for  R $> R_C$ .
}
\end{eqnarray}

 \begin{figure}[tp]
\centering
\includegraphics[scale=0.7]{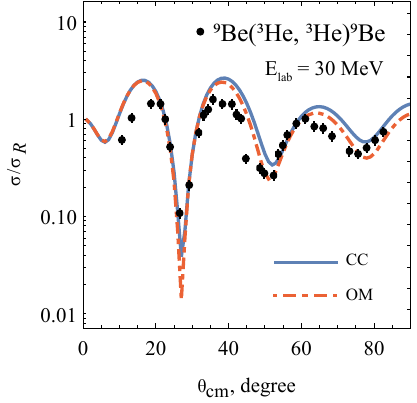}~~~~
\includegraphics[scale=0.7]{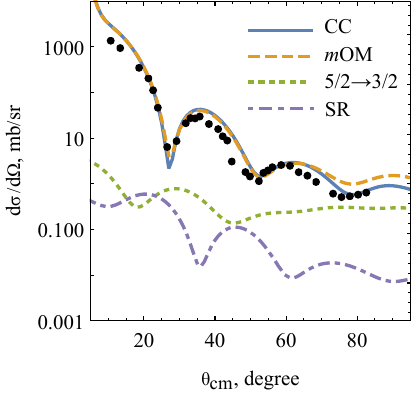}
\caption{  Theoretical cross sections for the elastic scattering  $^9{\rm Be}(^3{\rm He},~^3{\rm He})^9{\rm Be}$ in comparison with measured cross section data at ${\rm E}_{lab}=$30 MeV. Left panel: Calculated cross sections are given in ratio to the Rutherford scattering and carried out within the  coupled channels (CC, ${\chi}^2$ = 28.3) and optical model (OM, ${\chi}^2$ = 14.85) methods.
Right panel: 
 the same CC cross section in absolute unit, but presented in terms of the optical model cross section with modified potential parameters (mOM), the contribution of inelastic channel ($5/2 \rightarrow 3/2$) and the spin re-orientation effect (SR).
   
}
\label{elastic}
\label{pot}
\end{figure}

\begin{table*}[bp]
\footnotesize
\caption{\label{potpar} Potential parameters used in the optical model and CRC calculations.}
\center
\begin{tabular}{@{}llllllllllll@{}}
\toprule
 & $V_0$,  & $r_V^{a)}$, & $a_V$, & $W_0$, & $r_W^{a)}$, & $a_W$, & $V_{i}$, & $r_{i}^{a)}$, & $a_{i}$, & $r_C^{a)}$,      \\ 
  & MeV  & fm & fm & MeV  & fm & fm & MeV & fm & fm & fm   \\ \midrule
 \hethree+\benine & 101.9   & 0.700   & 0.777 & 30.81 & 0.854 & 0.817 & 2.5$^{c)}$      & 0.708    & 0.720     & 0.767 \\
  &103.9$^{b)}$&~&~&23.81$^{b)}$&~&~&~&~&~& \\
\beseven + \hefive & 258.516 & 0.588 & 0.726  & 18.0 & 0.773 & 0.6  & 45.4$^{d)}$      & 0.566    & 0.843     & 0.734  \\
\lisix + \lisix & 114.0 & 0.64 & 0.859  & 45.6 & 0.831 & 0.807  &   ~   &  ~    & ~     & 0.649  \\ 
\liseven + \lifive & 114.0 & 0.606 & 0.853  & 38.448 & 0.82 & 0.809  &   ~   &  ~    & ~     & 0.588  \\
$\alpha$ + \beeight & 121.0 & 0.252 & 1.01  & 17.0 & 1.38 & 0.34  &   ~   &  ~    & ~     & 0.724  \\ 
\bottomrule
\end{tabular}\\
\scriptsize
\flushleft
$^{a)}$ $r_i = R_i ( A^{1/3}_p+A^{1/3}_t )^{-1}$ \\
$^{b)}$ modified optical potential ($mOP$) used in the CRC calculations. \\
$^{c)}$ $i$ is for $SO$  \\
$^{d)}$ $i$ is for imaginary first derivative of volume potential  \\
\end{table*}

As a staring point for the seeking optical potential, we have taken global optical parameters from Ref. \cite{becchetti1971general}.
The elastic scattering cross section was fitted on the measured experimental data. It was performed by means of the \texttt{SFRESCO} \cite{thompson1988coupled}.
Obtained potential parameters are listed in Tab. \ref{potpar}.

Since the spin reorientation is not forbidden, e.g. $3/2 \rightarrow 3/2$ or $5/2 \rightarrow 5/2$, it was also taken into account in the CC calculations.
The calculated CC results are presented in Fig. \ref{elastic}.
The OM and CC calculations are almost identical. 
Explicit taking into account the coupling between inelastic channel and the coupling for spin reorientation modified optical potential. The real part of the optical potential becomes deeper, while
 the depth  of the imaginary part shortens (see mOP, Tab. \ref{potpar}).

The role of the inelastic channel, $5/2 \rightarrow 3/2$, in the elastic channel is turned out to be non-negligible. 
The effect of spin reorientation stands one magnitude lower than the OM estimation. 
The extrema point of SR coincides with the extrema of mOP.
This is one of the regular character of spin reorientation effects. Similar  behaviour was also reported in Ref. \cite{rudchik1996strong}.

\subsection*{Inelastic scattering}
The differential cross sections for the inelastic scattering of \hethree on \benine with the excitation 2.43 MeV is calculated in the framework of Coupled channels (CC) method \cite{thompson1988coupled}.
For the CC calculations, the mOP potential was used.

For the purpose of the explicit tracking of the couplings, we have not used a rotational model, but utilised the general case. In this case,  the strength of the coupling factor is given by:
\begin{equation}
RDEF(\lambda, J \rightarrow J') = (-1)^{J+J'+|J-J'|}
\sqrt{2J+1} \langle JK \lambda 0  \vert J' K \rangle \delta^{J\rightarrow J'}_\lambda 
\label{copling_factor}
\end{equation}
where $J$ and $J'$ are the spins of initial and final states, $K$ is the projection. The deformation length may have the form:
\begin{equation}
\delta^{J\rightarrow J'}_\lambda  = \beta^{J\rightarrow J'}_\lambda R_i.
\label{def_parameter}
\end{equation}
Here, $\beta^{J\rightarrow J'}_\lambda$ is the deformation parameters, $R_i$ is the interaction radius.

 \begin{figure}
\centering
\includegraphics[scale=0.7]{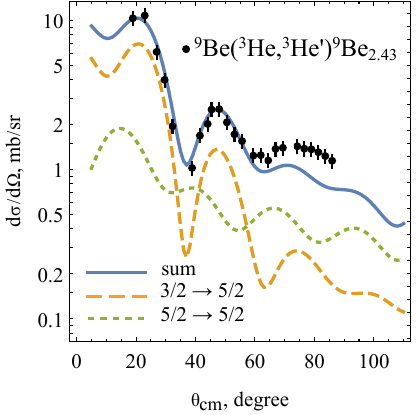}
\caption{   Inelastic scattering data in comparison with the CC calculation results in terms of  different coupling contributions: the transition from $\frac{3}{2}^-$ to $\frac{5}{2}^-$ states (dashed), and spin re-orientation for the state $\frac{5}{2}^-$ (doted). 
}
\label{inelastic}
\end{figure}

 The calculated differential cross sections for the inelastic scattering channel with \benine excitation at 2.43 MeV are shown in Fig. \ref{inelastic}.
 The spin re-orientation, i.e. the transition $\frac{5}{2} \rightarrow \frac{5}{2}$, has also been implemented in the coupling scheme.

The cross section of the excitation process underestimates the experimental data. However, the spin re-orientation effect enhances the cross section. Taken together, both processes are in good agreement with the experimental data. The cross section in the range starting from 65 degree is in exception, probably, due to the contributions of other excitation modes that were not included explicitly in the CC calculation.

In this study, the best agreement of theoretical analyses with data is obtained if one uses the deformation length $\delta^{3/2\rightarrow 5/2}_2=$1.97 fm. 
The study dedicated to the scattering of $\alpha$-particles on \benine from Ref. \cite{harakeh1980strong} report that the length is 1.61 fm, on the contrary the deformation length is 2.63 fm obtained in Ref. \cite{votava1973proton} in the reactions $p$ + \benine. 
Our taken result on the deformation length have the middle of the values in comparing with the presented other sources.  
In general, the structure of \benine shouldn't depend on the dynamics of nuclear reactions. Nevertheless, it can be explained by the different character of interactions of the projectiles with protons and neutrons in the target nucleus \cite{Jiang2020prob}. Therefore, the deformation lengths deduced directly from the inelastic scattering induced by different projectiles also might be various.

In terms of the Eq.(\ref{def_parameter}) the values of deformation parameters differs, since the interaction radius is ambiguous. If taken the radius parameter of the optical potential mOP $r=0.7$ fm, the deformation parameter has 0.8, which reproduces the previous study in Ref.\cite{janseitov2018investigation} with the same experimental data and applied theoretical method.

\section{Cluster-transfer reactions}

\subsection{\beseven + \hefive channel}

 \begin{figure}[tp]
\centering
\includegraphics[scale=1]{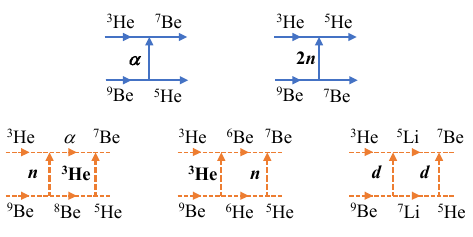}
\caption{
Reaction schemes for transfer mechanisms in \benine(\hethree, \beseven)\hefive: one-step (solid) and two-step (dashed) mechanisms.
}
\label{s_7be5he}
\end{figure}

 \begin{figure}[bp]
\centering
\includegraphics[scale=0.85]{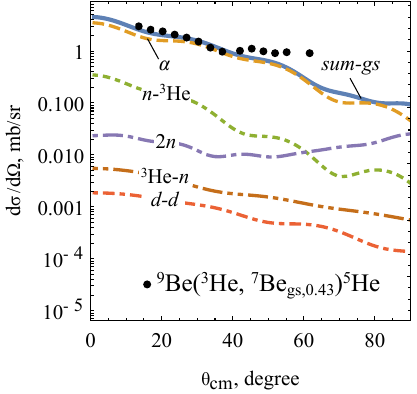}~~~~
\includegraphics[scale=0.85]{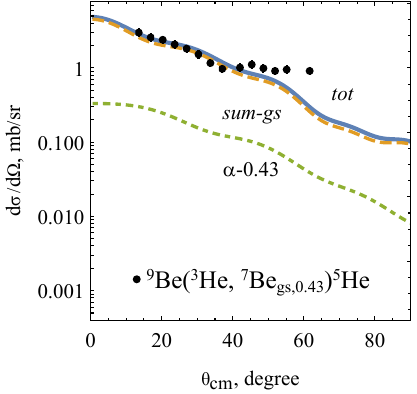}
\caption{
Calculated CRC cross sections with the experimental data for the nuclear reaction \benine(\hethree,$\rm^7Be_{gs, 0.43}$)\hefive.
 Cross sections are shown in terms of each considered transfer mechanisms to the ground state of \beseven (left panel). 
 Incoherent sum of cross sections over the ground and first excited 0.43 states of \beseven (right panel).
}
\label{7be5he}
\end{figure}

In this work the reaction \benine(\hethree, \beseven)\hefive is supposed to have both one-step and two-step transfer mechanisms which are demonstrated schematically   in Fig. \ref{s_7be5he}. 
The one-step mechanism occurs via transferring $\alpha$-particles, while the two-step mechanisms may have  $n$-\hethree, \hethree-$n$, $d$-$d$.
Moreover, we find that the transfer of $2n$ cluster is also possible at the backward angles of scattering.
Therefore, the CRC calculation covers all mentioned transfer mechanisms, and the resulting differential cross section for the reaction \benine(\hethree, \beseven)\hefive  is represented as
\begin{equation}
\frac{\rmd \sigma\left( \theta \right)}{\rmd \Omega} = 
\frac{1}{(2J_a+1)(2J_A+1)} 
~\vert 
f_{\rm I}\left( \theta \right)+
f_{\rm II}\left( \theta \right)
\vert^2,
\label{famplitudes}
\end{equation}
where the amplitudes of the one-step $f_{\rm I}\left( \theta \right)$ and two-step $f_{\rm II}\left( \theta \right)$ transfer mechanisms are given as follows
\begin{eqnarray}
f_{\rm I}\left( \theta \right)=& f_{\alpha}\left( \theta \right)+f_{2n}\left( \theta - \pi \right), \nonumber \\
f_{\rm II}\left( \theta \right)=& f_{n{\rm-^3He}}\left( \theta \right)+
f_{{\rm^3He-}n}\left( \theta \right)+
f_{d{\rm-}d}\left( \theta \right).
\end{eqnarray}
The optical potential mOP is chosen as the potential for the entrance channel. 
For the exit channels, we have utilized an optical potential with the global optical parameters from Ref. \cite{cook1982global}.

\begin{figure}[tp]
\centering
\includegraphics[scale=1]{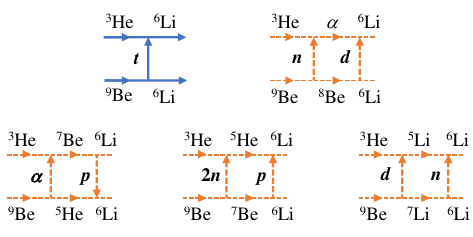}
\caption{
Same as the caption of Fig. \ref{s_7be5he} but for the reaction \benine(\hethree,~\lisix)\lisix.
}
\label{s_6li6li}
\end{figure}

\begin{figure}[bp]
\centering
\includegraphics[scale=0.85]{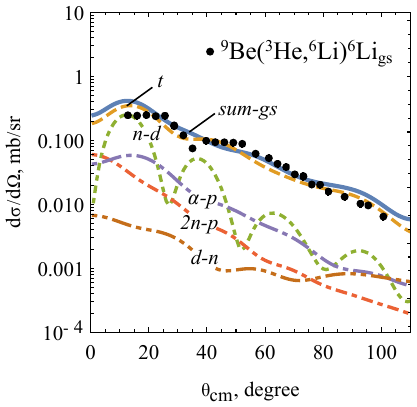}~~~~
\includegraphics[scale=0.85]{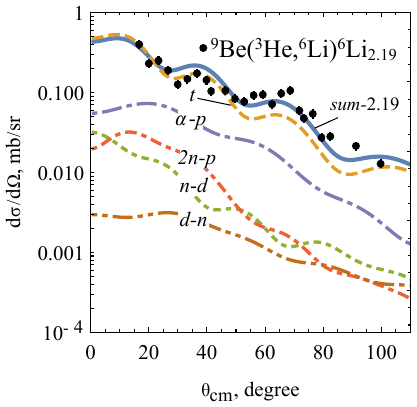}
\caption{  Experimental data for nuclear reaction $^9$Be($^3$He, $^6$Li)$^6$Li$_{gs}$ (left panel) and $^9$Be($^3$He, $^6$Li)$^6$Li$_{2.19}$ (right panel) compared to the calculated CRC cross sections in terms of channel contributions (For more, please, see the text).}
\label{6li6li0}
\end{figure}

The optical potential of the intermediate channel $\alpha$ + \beeight has been preferred to a potential, which reproduces the experimental cross sections of \benine(\hethree, $\alpha$)\beeight in the exit channel. 
The bound state wave functions have been built on the Woods-Saxon shaped potential, with the depths fitted on the binding energies. 
However, the resonance states were taken as the quasi-bound states, i.e. taken by means of the same procedure, but with the binding energy of 0.01 MeV. 
Spectroscopic amplitudes were taken from Refs. \cite{rudchik1996strong, kurath1975three, umbelino2019two}, and presented in Tab. \ref{tab:sa}.

All CRC calculations have been performed by means of the \texttt{FRESCO} code \cite{thompson1988coupled}. The two-step transfer reactions have been calculated by means of the $N$-step DWBA iteration neglecting the back couplings. The prior and post modes were used respectively for the first and second couplings, aimed at the avoiding the non-orthogonality term. The results of the CRC calculations are shown in Fig. \ref{7be5he}. The optical potentials used in these calculations are given in Tab. \ref{potpar}.

The direct transfer of $\alpha$-particle obviously prevails on other transfer mechanisms (see Fig. \ref{7be5he}, \textit{left panel}). 
It is interesting to note the different contribution of the transfer of the system $n$ + \hethree. 
It is turned out that the system $n$ + \hethree easily transferred in the way of $n$-\hethree rather that \hethree-$n$.
The reason of this may lie in the $p$-shell valence neutron, that loosely bound to \benine. 
Next contributor to the cross section is 2$n$ transfer at the backward scattering angles. 
Starting from 60$^\circ$ it gives up only  $\alpha$-transfer. 
However, starting from 90$^\circ$ the transfer mechanism 2$n$ may stand as a main contributor. 
To confirm this hypothesis, one must carry out the experiment at energies much higher than the energy in the present work, i.e. 30 MeV.
Thus, it would come to possible in the identifying of registered particles at the back hemisphere.
In similar  experimental studies \cite{umbelino2019two, rudchik1996strong}  the 2$n$ cluster transfer from \benine was also explored.
In particular, the transfer of 2$n$ cluster was observed by Umbelino \etal \cite{umbelino2019two} in the elastic transfer reactions \beseven + \benine.
The transfer mechanism $d$-$d$ has the lowest contribution among all of the proposed transfer mechanisms.

As a consequence of low detector resolution, we couldn't differ the registered  \beseven at the ground and first excited states.
Nevertheless, we could have estimated its weight to the cross section via CRC calculations.
It turned out the cross section to the first excited state impacts less notably than the transfer mechanism $n$-\hethree (see Fig. \ref{7be5he}, \textit{right panel}).
The calculated CRC cross sections reproduce experimental data well, except last few points, that have not been described with the proposed model.

\subsection{\lisix + \lisix channel}

\begin{figure}
\centering
\includegraphics[scale=0.85]{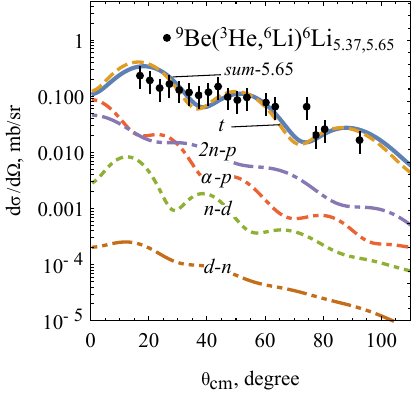}~~~~
\includegraphics[scale=0.85]{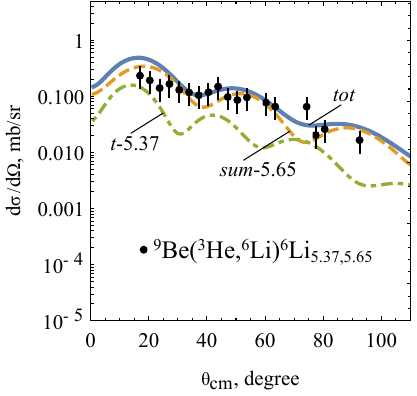}
\caption{   Experimental data for nuclear reaction \benine(\hethree, \lisix)\lisix$_{5.37, 5.65}$ compared to the  CRC calculation results.
The CRC cross section for the target-like nucleus $\rm{\lisix}^*$ at 5.65 MeV with the channel contributions (left panel). Total cross section summed incoherently  over 5.65 MeV and 5.37 MeV channels (right panel).

}
\label{6li6li2}
\end{figure}

The transfer mechanisms of the cluster $t$ is suggested to have also two-step processes as well, as it was suggested in Ref. \cite{rudchik1996strong}.
The direct transfer $t$, and two-step transfer mechanisms of the cluster $t$ (see Fig.~\ref{s_6li6li}): $n$-$d$, $\alpha$-$p$, 2$n$-$p$, and $d$-$n$, were included in the CRC calculations. Differential cross sections for these transfer mechanisms are obtained analogously as in Eq. \ref{famplitudes}.

CRC calculations for the transfer of the cluster $t$ have included the optical potential WS as for the entrance channel, and a Woods-Saxon shaped potential for the exit channel \lisix + \lisix. The potential parameters of the exit channel was extracted by fitting the experimental data on the elastic scattering \lisix + \lisix at the energy 40 MeV \cite{potthast1997global}. Obtained optical potential is shown in Tab. \ref{potpar}. 

Apart from the fact that \lisix was detected at the ground state, there were also two registered resonances. In the CRC calculations, they were taken into account through the coupling with the ground state of \lisix.

The calculated differential cross sections for the channel \lisix + \lisix at the ground state are presented in Fig. \ref{6li6li0} (\textit{left panel}) with the contributions of each transfer mechanisms.
The direct transfer of the cluster $t$ dominates at the whole range of angles. 
The two-step transfer mechanism $n$-$d$ is turned out to be less contributor than the transfer mechanism $t$ in contrary to the studies by Rudchik \etal \cite{rudchik1996strong} within the same reaction, but at the laboratory energy of 63 MeV.
In the work the role of the transfer mechanism $n$-$d$ prevails over other processes including the direct transfer $t$.
The cross section of the mechanism $n$-$d$ has an oscillatory character, and competes with the mechanism $\alpha$-$p$.
The latter transfer mechanisms, i.e. 2$n$-$p$ and $d$-$n$, are insignificantly involved to the reaction.

Contribution order changes, when \lisix gets excited (see Fig. \ref{6li6li0}, \textit{right panel}). 
The transfer mechanism $t$ remains unchanged, while $\alpha$-$p$ stands as the next contributor.
Weakening of the mechanism $n$-$d$ can be interpreted as the switching value in the spectroscopic amplitudes. 
In particular, the amplitude for the overlap $\langle ^6$Li$_{2.19}|^8$Be$\rangle$ with the configuration $2S_1$ is changed to 0.089. 
Such the small spectroscopic amplitude must have brought $n$-$d$ far below than the transfer mechanism $d$-$n$. However, the spectroscopic amplitudes with $1D_{1,2,3}$ have non-negligible values: 0.414, $-$0.477, and 0.744, that do not allow $n$-$d$ to disappear in the reaction.

Two resonances at 5.37 MeV and 5.65 MeV are mixed in the differential cross sections due to the experimental limitations. Therefore, we have included two channels into the CRC calculations.
Calculation results for \benine(\hethree, \lisix)\lisix$^*$ with the excitation 5.65 MeV are demonstrated in Fig. \ref{6li6li2} (\textit{left panel}) in comparing with experimental data.
The reaction mainly occurs through the direct transfer of cluster $t$.
Two transfer mechanisms, 2$n$-$p$ and $\alpha$-$p$, affect nuclear reaction only up to 30$^\circ$.
Other transfer mechanisms, i.e. $n$-$d$ and $d$-$n$, almost do not influence the reaction.

Due to the fact that the transfer of cluster $t$ has been the leading contributor in the \lisix~+~\lisix channel, the channel with the excitation 5.37 MeV calculated only with the transfer of $t$.
The differential cross section of the channel of the 5.37 MeV excited state
 added incoherently with the channel at 5.65 MeV. 
Total cross section resulting from the CRC calculations are shown in Fig.\ref{6li6li2} (\textit{right panel}).
The channel $\rm{\lisix}~+~\rm{\lisix}^*$ with the excitation 5.37 MeV has got less contribution than the channel at 5.65 MeV.
Calculated differential cross sections reproduces experimental data well.

\begin{table}[]
\footnotesize
\caption{Spectroscopic information data used in the CRC calculations for Composite ($A$) consisted of Core ($C$) and Valence particle ($\nu$) with configuration $nlj$. Most of Spectroscopic amplitudes (SA) were taken from Refs. \cite{umbelino2019two, rudchik1996strong, kurath1975three}}
\label{tab:sa}
\begin{tabular}{@{}lllllllr|lllllllr@{}}
\toprule
$A$    &       & $C$    &       & $\nu$  &     &$nlj$  & SA     & $A$    &       & $C$    &       & $\nu$  &     &$nlj$  & SA     \\ 
\midrule
$^6$Li & 1     & $^3$He & 0.5   & $t $    & 0.5 & 2$S_{0.5}$ & $-0.943$ & $^9$Be & 1.5   & $^5$He & 1.5   & $\alpha$& 0   & 2$D_{2  }$ & $-0.530$ \\ 
$^6$Li & 3     & $^3$He & 0.5   & $t $    & 0.5 & 1$D_{2.5}$ & $-0.943$ & $^7$Be & 1.5   & $^6$Li & $1_1$ & $p$     & 0.5 & 1$P_{0.5}$ & $-0.657$ \\
$^6$Li & 1     & $^3$He & 0.5   & $t $    & 0.5 & 1$D_{1.5}$ & $0.943 $ & $^7$Be & 1.5   & $^6$Li & 3     & $p$     & 0.5 & 1$P_{1.5}$ & $0.738 $ \\
$^9$Be & 1.5   & $^6$Li & $1_1$ & $t $    & 0.5 & 2$P_{0.5}$ & $-0.192$ & $^7$Be & 1.5   & $^6$Li & $1_2$ & $p$     & 0.5 & 1$P_{0.5}$ & $0.147 $ \\
$^9$Be & 1.5   & $^6$Li & $1_1$ & $t $    & 0.5 & 2$P_{1.5}$ & $-0.215$ & $^7$Be & 1.5   & $^6$Li & $1_2$ & $p$     & 0.5 & 1$P_{1.5}$ & $-0.132$ \\
$^9$Be & 1.5   & $^6$Li & 3     & $t $    & 0.5 & 2$P_{1.5}$ & $-0.594$ & $^7$Be & 1.5   & $^6$Li & $1_1$ & $p$     & 0.5 & 1$P_{1.5}$ & $-0.735$ \\
$^9$Be & 1.5   & $^6$Li & 3     & $t $    & 0.5 & 1$F_{3.5}$ & $-0.316$ & $^6$Li & $1_1$ & $^5$He & 1.5   & $p$     & 0.5 & 1$P_{0.5}$ & $-0.596$ \\
$^9$Be & 1.5   & $^6$Li & $1_2$ & $t $    & 0.5 & 2$P_{0.5}$ & $-0.118$ & $^6$Li & $1_1$ & $^5$He & 1.5   & $p$     & 0.5 & 1$P_{1.5}$ & $0.667 $ \\
$^9$Be & 1.5   & $^6$Li & $1_2$ & $t $    & 0.5 & 1$F_{2.5}$ & $-0.324$ & $^6$Li & 3     & $^5$He & 1.5   & $p$     & 0.5 & 1$P_{1.5}$ & $0.500 $ \\
$^4$He & 0     & $^3$He & 0.5   & $n $    & 0.5 & 1$S_{0.5}$ & $-0.741$ & $^6$Li & $1_2$ & $^5$He & 1.5   & $p$     & 0.5 & 1$P_{0.5}$ & $0.333 $ \\
$^9$Be & 1.5   & $^8$Be & 0     & $n $    & 0.5 & 1$P_{1.5}$ & $0.791 $ & $^6$Li & $1_2$ & $^5$He & 1.5   & $p$     & 0.5 & 1$P_{1.5}$ & $0.298 $ \\
$^9$Be & 2.5   & $^8$Be & 0     & $n $    & 0.5 & 1$P_{1.5}$ & $-0.816$ & $^5$Li & 1.5   & $^3$He & 0.5   & $d$     & 1   & 1$P_{1  }$ & $0.456 $ \\
$^9$Be & 2.5   & $^8$Be & 2     & $n $    & 0.5 & 1$P_{1.5}$ & $-0.986$ & $^5$Li & 1.5   & $^3$He & 0.5   & $d$     & 1   & 1$P_{2  }$ & $1.021 $ \\
$^9$Be & 2.5   & $^8$Be & 0     & $n $    & 0.5 & 1$P_{0.5}$ & $0.242 $ & $^9$Be & 1.5   & $^7$Li & 1.5   & $d$     & 1   & 2$S_{1  }$ & $-0.226$ \\
$^9$Be & 2.5   & $^8$Be & 2     & $n $    & 0.5 & 1$P_{0.5}$ & $0.417 $ & $^9$Be & 1.5   & $^7$Li & 1.5   & $d$     & 1   & 1$D_{1  }$ & $0.111 $ \\
$^6$Li & 1.0   & $^4$He & 0     & $d $    & 1.0 & 2$S_{1  }$ & $1.061 $ & $^9$Be & 1.5   & $^7$Li & 1.5   & $d$     & 1   & 1$D_{3  }$ & $-0.624$ \\
$^6$Li & 3.0   & $^4$He & 0     & $d $    & 1.0 & 1$D_{3  }$ & $1.061 $ & $^6$Li & $1_1$ & $^5$Li & 1.5   & $n$     & 0.5 & 1$P_{0.5}$ & $0.597 $ \\
$^8$Be & 0     & $^6$Li & $1_1$ & $d $    & 1.0 & 2$S_{1  }$ & $1.146 $ & $^6$Li & $1_1$ & $^5$Li & 1.5   & $n$     & 0.5 & 1$P_{1.5}$ & $-0.667$ \\
$^8$Be & 0     & $^6$Li & $1_1$ & $d $    & 1.0 & 1$D_{1  }$ & $0.112 $ & $^6$Li & 3     & $^5$Li & 1.5   & $n$     & 0.5 & 1$P_{1.5}$ & $1.095 $ \\
$^8$Be & 0     & $^6$Li & 3     & $d $    & 1.0 & 2$S_{1  }$ & $0.089 $ & $^6$Li & $1_2$ & $^5$Li & 1.5   & $n$     & 0.5 & 1$P_{0.5}$ & $-0.333$ \\
$^8$Be & 0     & $^6$Li & 3     & $d $    & 1.0 & 1$D_{1  }$ & $0.414 $ & $^6$Li & $1_2$ & $^5$Li & 1.5   & $n$     & 0.5 & 1$P_{1.5}$ & $-0.298$ \\
$^8$Be & 0     & $^6$Li & 3     & $d $    & 1.0 & 1$D_{2  }$ & $-0.477$ & $^7$Li & 1.5   & $^6$Li & $1_1$ & $n$     & 0.5 & 1$P_{0.5}$ & $-0.657$ \\
$^8$Be & 0     & $^6$Li & 3     & $d $    & 1.0 & 1$D_{3  }$ & $0.744 $ & $^7$Li & 1.5   & $^6$Li & $1_1$ & $n$     & 0.5 & 1$P_{1.5}$ & $-0.735$ \\
$^8$Be & 0     & $^6$Li & 3     & $d $    & 1.0 & 2$S_{1  }$ & $0.850 $ & $^7$Li & 1.5   & $^6$Li & 3     & $n$     & 0.5 & 1$P_{1.5}$ & $0.738 $ \\
$^8$Be & 2     & $^6$Li & 3     & $d $    & 1.0 & 2$S_{1  }$ & $0.747 $ & $^7$Li & 1.5   & $^6$Li & $1_2$ & $n$     & 0.5 & 1$P_{0.5}$ & $0.147 $ \\
$^8$Be & 2     & $^6$Li & 3     & $d $    & 1.0 & 1$D_{1  }$ & $0.079 $ & $^7$Li & 1.5   & $^6$Li & $1_2$ & $n$     & 0.5 & 1$P_{1.5}$ & $-0.132$ \\
$^8$Be & 2     & $^6$Li & 3     & $d $    & 1.0 & 1$D_{2  }$ & $0.259 $ & $^6$He & 0     & $^5$He & 1.5   & $n$     & 0.5 & 1$P_{1.5}$ & $-0.867$ \\
$^8$Be & 2     & $^6$Li & 3     & $d $    & 1.0 & 1$D_{3  }$ & $0.538 $ & $^7$Be & 1.5   & $^6$Be & 0     & $n$     & 0.5 & 1$P_{1.5}$ & $-0.935$ \\
$^8$Be & 0     & $^6$Li & $1_2$ & $d $    & 1.0 & 2$S_{1  }$ & $-0.237$ & $^6$He & 0     & $^4$He & 0     & 2$n$     & 1.0 & 1$S_{0  }$ & $0.909 $ \\
$^8$Be & 0     & $^6$Li & $1_2$ & $d $    & 1.0 & 1$D_{1  }$ & $0.372 $ & $^8$Be & 0     & $^6$Be & 0     & $d$     & 1.0 & 1$S_{0  }$ & $-1.200$ \\
$^8$Be & 2     & $^6$Li & $1_2$ & $d $    & 1.0 & 2$S_{1  }$ & $0.371 $ & $^7$Be & 1.5   & $^4$He & 0     & $^3$He  & 0.5 & 2$P_{1.5}$ & $-1.091$ \\
$^8$Be & 2     & $^6$Li & $1_2$ & $d $    & 1.0 & 1$D_{1  }$ & $0.053 $ & $^8$Be & 0     & $^5$He & 1.5   & $^3$He  & 0.5 & 2$P_{1.5}$ & $-1.102$ \\
$^8$Be & 2     & $^6$Li & $1_2$ & $d $    & 1.0 & 1$D_{2  }$ & $-0.356$ & $^6$Be & 0     & $^3$He & 0.5   & $^3$He  & 0.5 & 2$S_{0.5}$ & $0.943 $ \\
$^8$Be & 2     & $^6$Li & $1_2$ & $d $    & 1.0 & 1$D_{3  }$ & $-0.130$ & $^9$Be & 1.5   & $^6$He & 0     & $^3$He  & 0.5 & 2$P_{1.5}$ & $-0.215$ \\
$^5$He & 1.5   & $^3$He & 0.5   & 2$n$    & 0   & 1$P_{1  }$ & $-0.913$ & $^7$Be & 1.5   & $^6$Be & 0     & $n$     & 0.5 & 1$P_{0.5}$ & $-1.091$ \\
$^9$Be & 1.5   & $^7$Be & 1.5   & 2$n$    & 0   & 2$S_{0  }$ & $0.247 $ & $^6$He & 0     & $^5$He & 1.5   & $n$     & 0.5 & 1$P_{1.5}$ & $-1.102$ \\
$^9$Be & 1.5   & $^7$Be & 1.5   & 2$n$    & 0   & 2$D_{2  }$ & $0.430 $ & $^7$Be & 1.5   & $^5$Li & 1.5   & $d$     & 1   & 2$S_{1  }$ & $-0.647$ \\
$^6$Li & $1_1$ & $^5$He & 1.5   & $p $    & 0.5 & 1$P_{0.5}$ & $-0.597$ & $^7$Be & 1.5   & $^5$Li & 1.5   & $d$     & 1   & 1$D_{1  }$ & $-0.121$ \\
$^6$Li & $1_1$ & $^5$He & 1.5   & $p $    & 0.5 & 1$P_{1.5}$ & $0.667 $ & $^7$Be & 1.5   & $^5$Li & 1.5   & $d$     & 1   & 1$D_{3  }$ & $0.647 $ \\
$^7$Be & 1.5   & $^6$Li & $1_1$ & $p $    & 0.5 & 1$P_{0.5}$ & $-0.657$ & $^7$Li & 1.5   & $^5$He & 1.5   & $d$     & 1   & 1$D_{3  }$ & $0.647 $ \\
$^7$Be & 1.5   & $^6$Li & $1_1$ & $p $    & 0.5 & 1$P_{1.5}$ & $-0.735$ & $^7$Li & 1.5   & $^5$He & 1.5   & $d$     & 1   & 2$S_{1  }$ & $-0.647$ \\
$^7$Be & 1.5   & $^6$Li & 2     & $p $    & 0.5 & 1$P_{1.5}$ & $-1.095$ & $^7$Li & 1.5   & $^5$He & 1.5   & $d$     & 1   & 1$D_{1  }$ & $-0.121$ \\
$^7$Be & 1.5   & $^6$Li & $1_2$ & $p $    & 0.5 & 1$P_{1.5}$ & $-0.632$ & $^7$Li & 1.5   & $^5$He & 1.5   & $d$     & 1   & 1$D_{3  }$ & $0.647 $ \\
$^7$Be & 1.5   & $^6$Li & $1_2$ & $p $    & 0.5 & 1$P_{0.5}$ & $-0.632$ & $^7$Be & 1.5   & $^3$He & 0.5   & $\alpha$& 0   & 2$P_{1  }$ & $-0.950$ \\
$^7$Be & 1.5   & $^3$He & 0.5   & $\alpha$& 0   & 2$P_{1  }$ & $1.091 $ & $^9$Be & 1.5   & $^5$He & 1.5   & $\alpha$& 0   & 3$S_{0  }$ & $-0.810$ \\
$^9$Be & 1.5   & $^5$He & 1.5   & $\alpha$& 0   & 3$S_{0  }$ & $-0.810$ & $^9$Be & 1.5   & $^5$He & 1.5   & $\alpha$& 0   & 2$D_{2  }$ & $-0.536$ \\
\bottomrule
\end{tabular}
\end{table}

\subsection{\hesix + \besix channel}

\begin{figure*}[tp]
\centering
\begin{minipage}[c]{.5\textwidth}
\centering
\includegraphics[scale=1.]{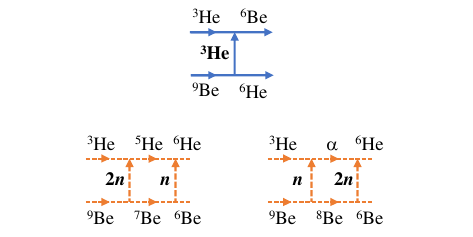}
\end{minipage}\hfil
\begin{minipage}[c]{.5\textwidth}
\centering
\includegraphics[width=7cm]{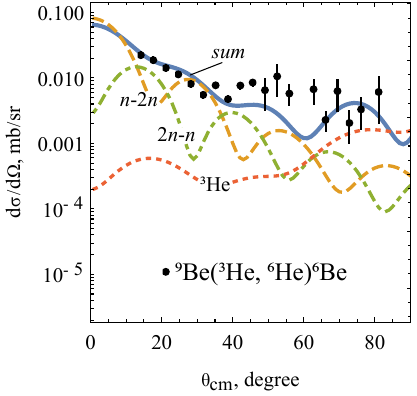}
\label{6he6be}
\end{minipage}
\caption{Left panel: reaction schemes for one- (solid) and two-step (dashed) transfer mechanisms for \benine(\hethree, \hesix)\besix. Right panel: comparison of the CRC differential cross sections with the experimental data for tha same reaction. 
\label{threenuclFig}
}
\end{figure*}


The channel \hesix + \besix is a unique channel, since the transfer of the system 3$n$ is observed at the forward angles of scattering. 
In his case, here we suggest three mechanisms of transfer (see Fig.~\ref{threenuclFig},~\textit{left panel} ): the two-step $n$-2$n$ and 2$n$-$n$, and the one-step transfer of the cluster \hethree at the back hemisphere. 
Differential cross sections are obtained as well as they are expressed in Eq. \ref{famplitudes}.
The resulting CRC calculations on cross sections in terms of each contributions followed three  mechanisms are demonstrated in Fig.~\ref{threenuclFig},~(\textit{right panel}).

The two-step transfer mechanisms, i.e. $n$-2$n$ and 2$n$-$n$,  compete with each other to give almost smooth cross section. 
Starting from 60$^\circ$ the most contribution is caused by the transfer of \hethree. 
Calculated CRC calculations are in good agreement with the experimental data.

 However the CRC calculations for the channel \hesix + \besix have been performed using the large valued spectroscopic amplitudes . For example, the overlaps $\langle$\hesix~$|$~$\alpha \rangle,~\langle$\besix~$|$~\beeight$\rangle$ have the amplitudes $\sim 1.2$.
On the contrary the studies  \cite{oganessian1999dynamics,khoa2004di} reported the amplitude for $\langle$\hesix~$|$~$\alpha \rangle$ equals $\sim 1.0$.
Such kind of difference between spectroscopic amplitudes points out that there may have other processes not included in the reaction model. 
Probably, the one-step transfer of the system 3$n$ can take  place in the channel \hesix + \besix. This problem is a separate topic of study, and it cannot precisely solved by the model shown in this work. For more accurate results one might apply using the four body problem.


\section*{Conclusion}
The nuclear reactions with \benine induced with the \hethree at 30 MeV was investigated. 
Elastic channel treated within the optical model, while inelastic channel was considered by means of the Coupled channels approach.
New parameters of the optical potential for the system \hethree + \benine at 30 MeV were obtained.
Using the optical potential the deformation parameter $\delta^{3/2\rightarrow 5/2}_2=0.8$ was extracted within the CC method, which could reproduce the results from another source \cite{janseitov2018investigation}. 

All possible transfer mechanisms were suggested in the cluster transfer channels, \beseven + \hefive, \lisix + \lisix, and \hesix + \besix.
Differential cross sections for these transfer mechanisms were calculated in the CRC framework using the Spectroscopic amplitudes and optical potentials without any corrections. 
The primary pick-up of valence neutron turned out to be easily picked-up in the cluster transfer channels, except in the channels with the \lisix excitations. 
This manner specifies the cluster structure $n$ + \beeight in \benine.

Special attention attracts the channel \hesix + \besix. In the CRC framework, we could obtain good agreement of the calculated cross sections with data suggesting the two-step transfer mechanisms, $n$-2$n$ and 2$n$-$n$. However, the large valued spectroscopic amplitudes used in the CRC point out that there may play other processes, e.g. direct transfer of three neutrons.

The one-nucleon transfer channels, as well as the charge exchange reaction channels, are planned to investigate as a continued of the series of studies dedicated to the reaction \hethree + \benine at the energy of 30 MeV.

\section*{Acknowledgements}
We would like to thank the NPI (Nuclear Physics Institute) \v{R}e\v{z} for giving us the opportunity to perform this study, as well as the cyclotron staff of institutes for the excellent beam quality.

This research was funded by the Science Committee of the Ministry of Science and Higher Education of the Republic of Kazakhstan (Grant No. AP14870958).

\section*{References}
\bibliographystyle{iopart-num}
\bibliography{bibliography}

\end{document}